\documentclass[aip,jcp,preprint,superscriptaddress,floatfix]{revtex4-1}
\usepackage{epsfig}
\usepackage{epsfig}
\usepackage{color}

\begin{document}
\title{A Transition State Theory for Calculating Hopping Times and Diffusion in Highly Confined Fluids.}

\author{Surajith N. Wanasundara}
\affiliation{Department of Chemistry, University of Saskatchewan, Saskatoon, SK, S7N 5C9, Canada}

\author{Raymond J. Spiteri}
\affiliation{Department of Computer Science, University of Saskatchewan, Saskatoon, SK, S7N 5C9, Canada}

\author{Richard K. Bowles}
\email{richard.bowles@usask.ca}
\affiliation{Department of Chemistry, University of Saskatchewan, Saskatoon, SK, S7N 5C9, Canada}

\date{\today}

\begin{abstract}
Monte Carlo simulation is used to study the dynamical crossover from single file diffusion to normal diffusion in fluids confined to narrow channels. We show that the long time diffusion coefficients for a series of systems involving hard and soft interaction potentials can be described in terms of a hopping time that measures the time it takes for a particle to escape the cage formed by its neighbors in the pore. Free energy barriers for the particle hopping process are calculated and used to show that transition state theory effectively describes the hopping time for all the systems studied, over a range of pore diameters. Our work suggests that the combination of hopping times and transition state theory offers a useful and general framework to describe the dynamics of these highly confined fluids.

\end{abstract}
%Keywords: single file diffusion, Fickian diffusion, free energy barrier, Monte Carlo simulation

\maketitle

\newpage
\section{Introduction}

Confined particles exhibit dynamical properties that are very different from those observed in bulk fluids.~\cite{Gelb:1999p1920,Gubbins:2010p11770} In particular, if the particles are confined to long narrow pores such that they are unable to pass each other and are subject to a Brownian background~\cite{Levitt:1973p14144} or stochastic forces~\cite{Percus:1974wp}, they exhibit a form of anomalous diffusion where the mean squared displacement (MSD) of the particles along the pore axis increases as the square root of time ($t^{1/2}$). In the long time limit, the MSD can be described by an Einstein-like relation,
\begin{equation}
  \langle \Delta x^2 (t) \rangle = \langle (x(t)-x(0))^2\rangle = 2F_xt^{1/2}\mbox{ ,}\\
 \label{eq:ad} 
\end{equation}
where $F_x$ is the mobility factor of the particles along the axial direction of the channel and the $\langle \cdots \rangle$ denotes the average over the particles. This dynamic phenomenon, known as single file diffusion (SFD), has been observed in the transport of fluids through zeolites,~\cite{Karger1992,Gupta1995,Kukla1996,Hahn1996} carbon nanotubes,~\cite{Mukherjee2010,Chen:2010p11420} and the channels formed in some metal organic frameworks.~\cite{Salles:2011p15709} SFD has also been observed in the diffusion of colloidal particles along narrow channels~\cite{Wei:2000ei,Lin:2005p15430} and has been used to provide time-controlled drug delivery.~\cite{Yang:2010p15731}

When the diameter of the channel is just wide enough so that particles can pass each other, they are no longer permanently caged by their neighbors and the system exhibits a dynamical crossover from SFD  to normal, Fickian diffusion in the long time limit, with a MSD given by
\begin{equation}
  \langle \Delta x^2 (t) \rangle= 2D_xt\mbox{ ,}\\
  \label{eq:nd}
\end{equation}
where $D_x$ is the diffusion coefficient along the $x$ direction. Understanding the nature of this crossover regime holds the key to developing diffusion control in nanofluidic devices and using porous materials for the separation of different size particles. For example, it is possible to establish dual mode diffusion~\cite{Sholl1997,Hahn:1998hh,Sholl:1998ws,Adhangale2002,Adhangale:2003p15701} in particle mixtures, where the large particles diffuse anomalously and the small particles diffuse normally, by selecting a channel diameter that is narrow enough to prevent the large particles from passing while allowing the smaller particles through. Recent simulation studies,~\cite{Ball:2009p6393,Wanasundara2012} identifying optimal pore diameters that maximize the difference in the diffusion rates of the dual mode components, suggest the method may be ideal for challenging separations involving components that are chemically similar and only marginally different in size.

The dynamics of the particles in the crossover regime can be characterized in terms of a hopping time, $\tau_{hop}$, that measures the average time it takes for a particle to escape the cage formed by its two neighbors.~\cite{Mon:2002p4642} For time intervals shorter than $\tau_{hop}$, particles undergo anomalous diffusion and the MSD is proportional to $t^{1/2}$. After $\tau_{hop}$, a particle can hop past one of its neighbors, in either direction, where it again becomes temporarily trapped. If both the units of time and distance are rescaled by $\tau_{hop}$ and the distance travelled in $\tau_{hop}$ respectively, then this hopping process leads asymptotically to normal diffusion in the long time limit with a diffusion coefficient that varies as
\begin{equation}
D_x \sim 1/\tau^{1/2}_{hop}\mbox{ .}
\label{eq:dxhop}
\end{equation}

An appealing feature of the hopping time approach is that the long time diffusive behavior of the fluid is captured by a single, local parameter that contains all the important details of the system such as the particle-particle interactions, the particle-wall interactions, and the density. Furthermore, $\tau_{hop}$ can also be measured in simulations and is accessible to theoretical analysis. Simulations of confined hard spheres in two~\cite{Mon:2007p2259,Ball:2009p6393} and three~\cite{Mon:2002p4642} dimensions verified Eq.~\ref{eq:dxhop} and showed that the hopping time behaves as a power law in the reduced pore diameter, $R_p-\sigma$, so that 
\begin{equation}
\tau_{hop}\sim(R_P - \sigma)^{-\alpha}\mbox{,}\\
\label{eq:hoprp}
\end{equation}
where $R_p$ is the channel radius and $\sigma$ is the hard particle diameter. $\tau_{hop}$ diverges at the passing threshold, $R_p=\sigma$, where the hard particles are no longer able to pass each other and become permanently caged. The exponent $\alpha$ was found to be equal to the dimensionality of the system, $d$. Mon and Percus~\cite{Mon:2006p2267} used a finite difference method to solve the multidimensional diffusion equation of two hard disks in a narrow flat channel with hard walls and found that $\alpha = 2$, in agreement with a simple transition state theory (TST) that examines the rate at which two hard particles in a channel can pass each other.~\cite{Bowles:2004p7}  However, approaches using dimensionally reduced Fick-Jacobs schemes,~\cite{Zwanzig:1992ig} involving the projection of the diffusion equation of the probability density in higher dimensions onto one dimension, generally find $\alpha=d-1/2$.~\cite{Kalinay:2007p2381} Why the two theoretical methods disagree remains unclear, but recent Brownian dynamics simulations~\cite{Mon:2008wv,Mon2009} of a simplified model for the hard particle passing problem found the TST result was correct only when the channel radius was extremely narrow and suggested that the assumptions involved in the projection techniques may no longer be valid in a region where the reduced channel geometry is varying rapidly, i.e., near the passing threshold.

To date, hard particle systems with hard walls have been the focus for studies of the relationship between the hopping time and the diffusion coefficient because there is a well defined passing threshold. There is no such passing threshold for particles interacting with each other or the walls with soft potentials, and such systems should always exhibit normal diffusion in the long time limit at all channel diameters. However, if the channel is sufficiently narrow, the particles will need to overcome a free energy barrier in order to pass each other. When this free energy barrier is high, passing events will become rare and we would still expect to see SFD for long periods of time before the crossover to normal diffusion occurs. Measurements of SFD transport in zeolites by molecular dynamics simulations, using realistic potentials, have exhibited a long time crossover to normal diffusion.~\cite{Hahn1998, Tepper1999} Furthermore, a recent Brownian dynamics simulation of the SFD-normal diffusion crossover in a colloidal particle system, modelled using very steep, $(1/r_{ij})^{48}$ interaction potentials, where $r_{ij}$ is the separation between particles, found $\tau_{hop}$ still followed the TST power law scaling in the reduced pore diameter and Eq.~\ref{eq:dxhop} remained valid.~\cite{Sane2010}

% need to frame this in context of liquids more clearly.

The goal of the present work is to show that the hopping time approach, combined with TST, has the potential to provide a predictive tool for understanding the long time diffusive properties of confined fluids in the SFD-normal diffusion crossover regime for a wide class of systems.  We show that Eq.~\ref{eq:dxhop} is still valid for systems with soft repulsive interactions $(1/r_{ij})^{n}$, $n=6$, 12, and for particles interacting through the Lennard-Jones potential. We also develop a TST approach to finding $\tau_{hop}$ that is widely applicable to most systems of interest by focusing on calculations of the free energy barrier associated with two particles passing. Our approach is  found to remain valid even for channel diameters and systems where the power law scaling laws in terms of the reduced channel radius is no longer applicable, highlighting its generality. 

%The remainder of the paper is organized as follows: The application of TST to the calculation of hopping times in single file diffusion is developed in Section 2. Section 3  outlines the models studied and our simulation methods. Section 5 describes our results and discussion. Our conclusions are contained in Section 5.

\section{Transition State Theory for Hopping Times}

% barrier crossing approaches, 
The problem of measuring the time (or rate) of escape from a metastable state appears in almost every area of science and is the focus of a diverse range of fields including the study of chemical kinetics, diffusion, and nucleation theory.~\cite{Hanggi:1990en} For the classic problem of a Brownian particle diffusing over an external potential energy barrier, the escape time can be obtained using the approaches developed by Smoluchovski~\cite{smol1917} and Kramer~\cite{kramers1940} in limits of high and low friction, respectively. Transition State Theory, originally developed by Eyring~\cite{eyring1935} in form of activated rate theory, is an equilibrium based approach that neglects the effects of barrier recrossing. As such, it provides an upper bound to the rate and is most appropriately applied in the Kramer turnover regime where friction is unimportant. However, TST also provides effective estimates of the rate even when the details of the frictional force is not known.  Bennent~\cite{Bennett1975} and Chandler~\cite{Chandler:1978hh} developed a general approach to barrier crossing applicable to cases where the barrier is high relative to the thermal energy in the system that reduces to TST when recrossing is not considered. The basic principle underlying all of these approaches is that the barrier crossing rate is proportional to the probability of finding the system at the top of the barrier and a number of computational techniques now exist for calculating this probability even when barrier crossing events are rare.~\cite{Frenkel2002} The time of escape is then inversely proportional to the rate.

%Transition state theory suggests that the time it takes for a system to escape a metastable state in an activated process is inversely proportional to the probability of finding the system in the transition state, which is then related exponentially to height of the free energy barrier associated with the transition. 

For the problem of particle hopping in single file diffusion, we are interested in calculating the time it takes for a particle to escape the cage formed by its two neighbors on either side. As two particles approach each other within the confines of the pore, the particle--particle exclusion leads to a restriction in configuration space, creating a free energy barrier with a transition state located at the point where the two particles are side by side in the plane perpendicular to the longitudinal axis of the pore. At low densities, the transition state should not be affected by the presence of the other particles, suggesting that a two particle description may be sufficient to capture the key elements of the activated particle hopping. We can then define a reaction coordinate for the hopping process as $\Delta x=x_2-x_1$, where $x_1$ and $x_2$ are the positions of the particles along the longitudinal axis of the pore so that the transition state is located at $\Delta x=0$.  Negative and positive values of $\Delta x$ represent the reactant and product states, respectively. 

The canonical partition function along the reaction coordinate for our two particle system can be expressed as
\begin{equation}
Q_R(\Delta x^{\prime})=\frac{1}{N!\Lambda^{3N}}\int e^{-\beta U(q)}\delta(\Delta x^{\prime}-\Delta x(q))dq\mbox{,}\\
\label{eq:qr}
\end{equation}
where $N$ is the number of particles, $\Lambda$ is the de Broglie wavelength, and $\beta=1/k_B T$, with $k_B$ being Boltzmann's constant and $T$ the temperature. $U(q)$ is the potential energy as a function of the particle coordinates, $q=(q_1,q_2)$. If we define the reaction coordinate dependent free energy as $F_R(\Delta x^{\prime})=-kT\ln Q_R(\Delta x^{\prime})$, TST suggests
\begin{equation}
\ln \tau_{hop}\propto \beta\Delta F\mbox{ ,}
\label{eq:lntaudf}
\end{equation}
where $\Delta F=F_R(0)-F_R(\infty)$. 

$\Delta F$ only contains integrals involving two particles and can be calculated analytically in many cases. This approach was used to obtain the TST power law dependence of $\tau_{hop}$ on $R_p-\sigma$ for two hard disks confined by hard walls.~\cite{Bowles:2004p7} However, solving the partition function, even for two particles, becomes difficult for the complex particle--particle and particle--wall interaction of more realistic systems and computer simulation free energy calculations offer a convenient alternative. Our goals are to demonstrate that Eq.~\ref{eq:lntaudf} holds for a variety of different systems, over a range of pore diameters, and show that the hopping time approach, combined with TST, offers a general method for studying the diffusion of confined fluids in the single file - normal diffusion crossover regime.

\section{Models and Methods}

\subsection{Models}

We study a range of three dimensional systems, each consisting of $N$ particles confined to a long, structureless, cylindrical pore of diameter, $D_p=2R_p$. The axial direction of a pore of length, $L$, extends along the $x$-axis of our coordinate system, and we use periodic boundary conditions in this direction.  Four particle-particle interactions are considered including hard spheres,
\begin{equation}
   U_{HS}(r_{ij}) = \left\{ \begin{array}{rl} 
 0 &\mbox{ if $r_{ij}  \geq \sigma$} \mbox{ ,}\\
 \infty &\mbox{ if $r_{ij} < \sigma$} \mbox{ ,}
       \end{array} \right. 
       \label{eq:vij}
\end{equation}
where $r_{ij} = |{\bf  r}_i - {\bf r}_j| $ is the distance between particles $i$ and $j$, two soft-sphere repulsive potentials,
\begin{equation}
 U_{SR_k}(r_{ij}) = \epsilon \left( \frac{\sigma}{r_{ij}} \right)^k\; ,\  k =6\mbox{ and }12\mbox{ ,}
 \label{repultive}
\end{equation}
and the Lennard--Jones (LJ) potential,
\begin{equation}
U_{LJ}(r_{ij}) = \epsilon\left[ \left( \frac{\sigma}{r_{ij}} \right)^{12} - \left( \frac{\sigma}{r_{ij}} \right)^6 \right]\ \mbox{ .} 
\end{equation}
Here, $\epsilon$ and $\sigma$ represent the energy and length interaction parameters, respectively.

The two particle-wall interactions studied are defined in terms of the radial position of the particle in the pore, $\hat{r}_i=(y_i^2+z_i^2)^{1/2}$, where $y_i$ and $z_i$ are the particle positions in the $yz$--plane relative to the center of the pore. The hard wall interaction is given by
\begin{equation}
U_{WHS}(\hat{r}_i) = \left\{ \begin{array}{rl}
 0 &\mbox{ if } \hat{r}_i \leq (D_P - \sigma)/2 \mbox{ ,}\\
 \infty &\mbox{ if } \hat{r}_i > (D_P - \sigma)/2 \mbox{ ,}
       \end{array} \right. 
       \label{eq:vwi}
\end{equation}
and the soft--repulsive wall interaction is 
\begin{equation}
 U_{WSR}(\hat{r}_{i}) = \epsilon \left( \frac{\sigma}{D_p/2-\hat{r}_{i}} \right)^{12}\mbox{ .}\\
\end{equation}
In this study, we report results for the five different systems with particle-particle : particle-wall interaction combinations, $(U_{HS}:U_{WHS})$,  $(U_{SR_{12}}:U_{WHS})$, $(U_{SR_6}:U_{WHS})$, $(U_{LJ}:U_{WHS})$  and $(U_{SR_{12}}:U_{WSR})$.

\subsection{Mean Squared Displacement and Average Hopping Time}

The MSD and $\tau_{hop}$ are calculated over a range of pore diameters in a series of canonical ($N,V,T$) simulations, with $N=5000$ and $T=1$ in units of $\epsilon k_B^{-1}$.  However, SFD only arises in structureless, linear tubes, such as those considered in the current work, when there is a random component to the dynamics. For example, SFD has been observed in fluids under the influence of a random background or stochastic forces. Molecular dynamics simulations using inertial dynamics simply leads to normal diffusion unless some degree of randomness is added to the dynamics explicity.~\cite{Hahn:1996ja,Mon:2003ed} To ensure that we observe SFD in our systems, we use Monte Carlo dynamics, which involves moving particles according to the standard Metropolis Monte Carlo (MC) algorithm~\cite{Frenkel2002} where, for each attempted MC move, we randomly select a particle and move it in a randomly selected direction within a range bounded by a maximum displacement of $0.05\sigma$. We define a unit of time as an MC cycle, consisting of $N$ MC attempted particle moves, and each simulation runs for $2 \times 10^8$ MC cycles. The simulation scheme used here follows that outlined in previous studies of hopping times in hard particle systems,\cite{Mon:2002p4642,Ball:2009p6393} and recent simulations studies~\cite{Sanz:2010if,Patti:2012uf} have shown that these types of Monte Carlo based dynamics provide a computationally efficient, coarse--grained approximation to Brownian motion in the absence of hydrodynamic interactions.

The MSD is calculated as
\begin{equation}
 \langle \Delta x^2(t) \rangle = \frac{1}{N} \sum_{i=1}^{N} (x_i(t) - x_i(0))^2\ ,
  \label{eq:msdsim}
\end{equation}
where $x_i(0)$ and $x_i(t)$ are the positions of particle $i$ along the pore axis at the time origin and a later time, $t$, respectively. Improved averaging of the MSD is obtained by using 20 independent time origins separated by $1 \times 10^7$ MC cycles.  The diffusion coefficient, $D_x$, is then calculated from the slope of the MSD plotted as a function of time, using data between $t = 4 \times 10^6$ and $1 \times 10^7$ MC steps, where it increases linearly as a function of $t$. One thousand MC cycles were performed before measurements of the MSD were started. The simulation scheme used here follows that outlined in previous studies of hopping times in hard particle systems.\cite{Mon:2002p4642,Ball:2009p6393}

The hopping time is the average number of MC cycles needed for a particle to hop over one of its nearest neighbor particles and is calculated as follows: After one thousand MC cycles are performed to bring the system to equilibrium, the hopping time for all the particles is initially set to zero. The hopping time for any given particle is then measured as the number of MC cycles it takes for the particle to pass one of the two immediate neighbors, that form its cage in the single file of particles. The time for the hopping event is recorded and the hopping time for the particle is reset to zero. The process is then repeated and the average hopping time is calculated over all the hopping events for all of the particles. We also performed some simulations with $N=9000$ particles to check system size effects. These tests confirm that 5000 particles are sufficient to calculate the hopping time and the diffusion coefficient.

To compare results from different pore diameters and different types of potentials, the appropriate linear density, $\rho_L=N/L$, for each state point studied is determined by performing standard constant pressure MC simulations~\cite{Frenkel2002} at $P_L\sigma^3/k_BT$ = 0.4, where $P_L$ is the longitudinal pressure imposed on the pore ends. The ($N,P_L,T$) simulations were carried out using $N=1000$ and $10^9$ MC cycles after equilibrium was established. The ranges of $D_p/\sigma$ and $\rho_L \sigma$ studied for our model systems are given in Table \ref{Tab:density}.

\subsection{Free Energy Barrier Calculations}

The free energy profile along the reaction coordinate, $F_R(\Delta x)$, is calculated using the umbrella sampling method.~\cite{Torrie1974,Frenkel2002} We use a harmonic biasing potential with a spring constant of 50~$k_BT$ and 27 independent simulation windows with umbrella centers spaced over the range $\Delta x/\sigma=[0,2]$. The initial particle positions in each window are set with $\Delta x/\sigma$ equal to the umbrella center for the window and the particles moved, subject to the biased potential acceptance rules, by randomly selecting one of the two particles and moving it in a randomly selected direction within a range bound by a maximum displacement of $0.05\sigma$. A total of 10$^9$ MC were sampled in each window and the complete, unbiased free energy profile is reconstructed from the combined simulations using the weighted histogram analysis method (WHAM).\cite{Kumar1992} Free energy profiles are calculated for all pore diameters considered for hopping time calculations.

\section{Results and Discussion}

\begin{figure}[ht]
\includegraphics[width=7.00in]{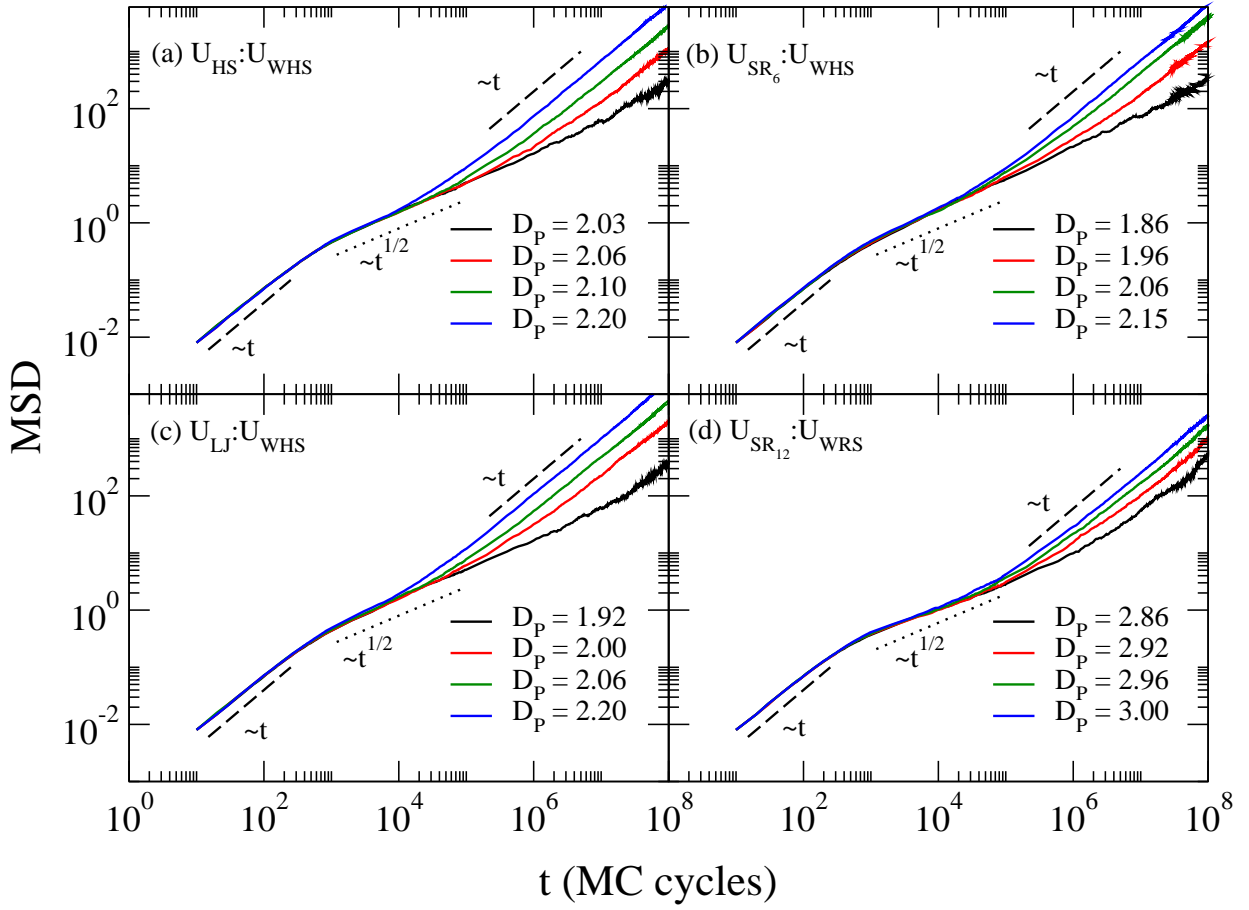}
\caption{The MSD as a function of time, $t$, for systems (a) $U_{HS}:U_{WHS}$ (b) $U_{SR_6}:U_{WHS}$ (c) $U_{LJ}:U_{WHS}$ and (d) $U_{SR_{12}}:U_{WSR}$ for a range of pore diameters. The dashed and dotted lines indicate line slopes  for $\sim t$ and $\sim t^{1/2}$, respectively, for comparison.}
 \label{MSD_mcsteps}
\end{figure}

Figure~\ref{MSD_mcsteps} plots the MSD as a function of time for four of the systems studied, for a variety of different pore diameters, and shows that all the systems exhibit the same three distinct diffusion regimes. The results for the $U_{SR_{12}}:U_{WHS}$ system that are not shown are qualitatively the same (see Supplementary Information). At short times, when the particles have moved over short distances relative to their diameter, the forces arising from particle-particle interactions effectively remain constant, and the MSD of the particles increases linearly with a diffusion coefficient that is characteristic of an isolated particle in a medium.~\cite{pusey:LFG1991} The MC dynamics employed in our simulations provide us with a computationally inexpensive imitation of Brownian motion and we do not see the quadratic behavior of the ballistic regime observed in deterministic type dynamics. At intermediate times, the particles interact with their caging neighbors but are unable to pass. This leads to SFD and the MSD increases as $t^{1/2}$. Eventually, the particles are able to pass at long times and the MSD returns to normal diffusion. These results confirm the presence of the SFD--normal diffusion crossover shown previously for hard potentials~\cite{Mon:2002p4642,Mon:2006p2267,Bowles:2004p7,Mon:2007p2259} as well as for the LJ potentials.~\cite{Hahn1998, Tepper1999} It is also important to note that the characteristic time it takes for the system to return to normal diffusion increases as the pore size decreases. At $D_p=2.0\sigma$, the $U_{HS}:U_{WHS}$ system reaches the passing threshold, where the particles become permanently caged by their neighbors, and the SFD-normal diffusion crossover would be suppressed for narrower channels. However, we see the crossover continues for soft particle systems for much smaller $D_p$. 

\begin{figure}[ht]
\includegraphics[width=7.00in]{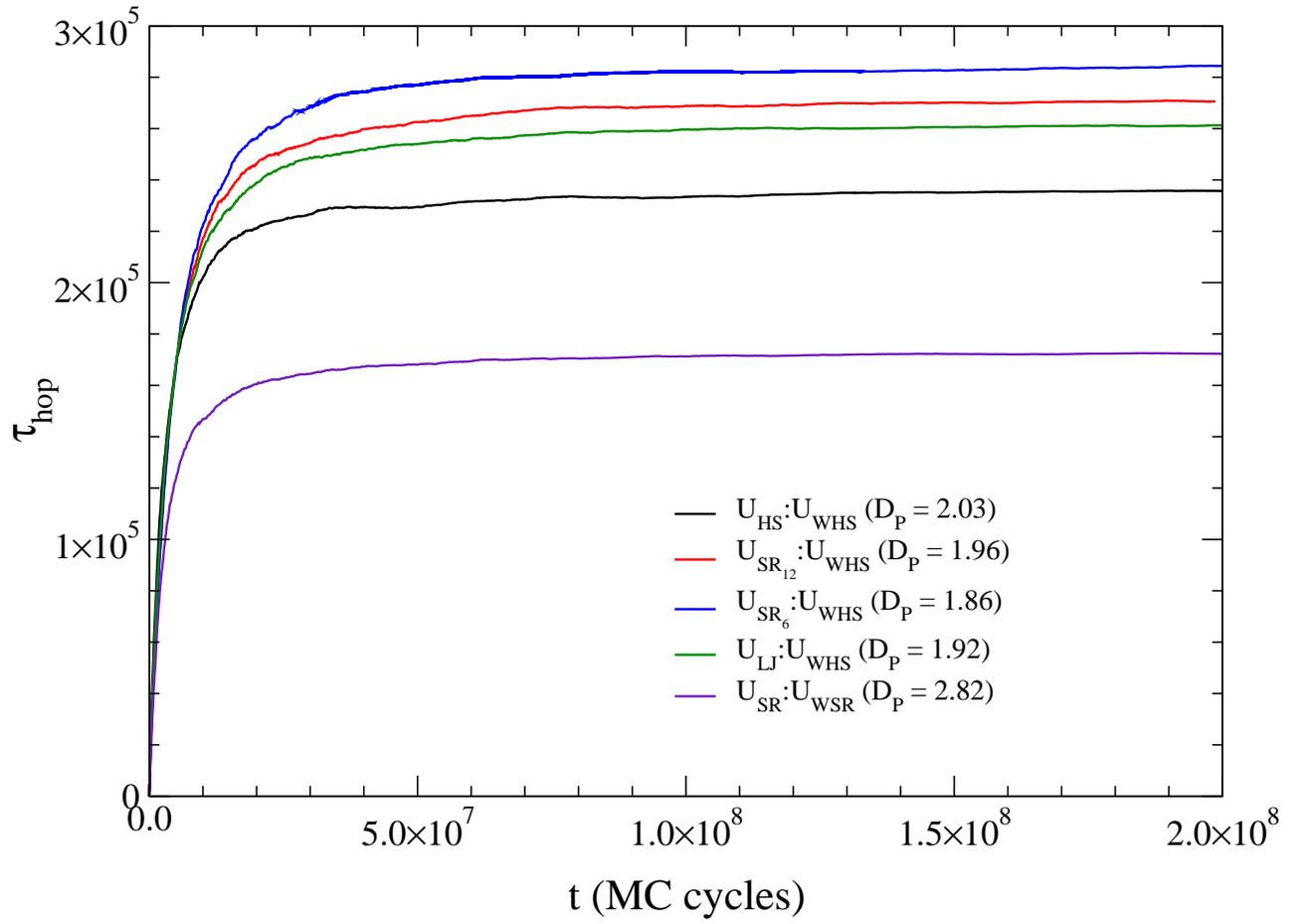}
\caption{The average hopping time, $\tau_{hop}$, as a function of time, $t$, for the narrowest pore diameter studied for each system.}
 \label{hopping}
\end{figure}

A previous study~\cite{Ball:2009p6393} found that $\tau_{hop}$ was slow to converge and extrapolation functions involving stretched exponentials were used to obtain accurate estimates. In Figure~\ref{hopping} we plot the average hopping time, calculated over all hopping the events that have occurred before time $t$, as a function of $t$ for the narrowest of pores studied for each system. When the pore diameter is small, hopping becomes more difficult so $\tau_{hop}$ becomes sensitive to rare hopping events that have large values and contribute significantly to the average. As a result, $\tau_{hop}$ is underestimated at short simulation times because the slow hopping particles have not yet contributed to the average and $\tau_{hop}$ only converges to its true value for simulation times much longer than these slow hopping times. We find that $\tau_{hop}$ has converged on the timescales of the long simulations used here and we use these values directly. $\tau_{hop}$ converges more quickly for the wider pores.

\begin{figure}[ht]
\includegraphics[width=7.00in]{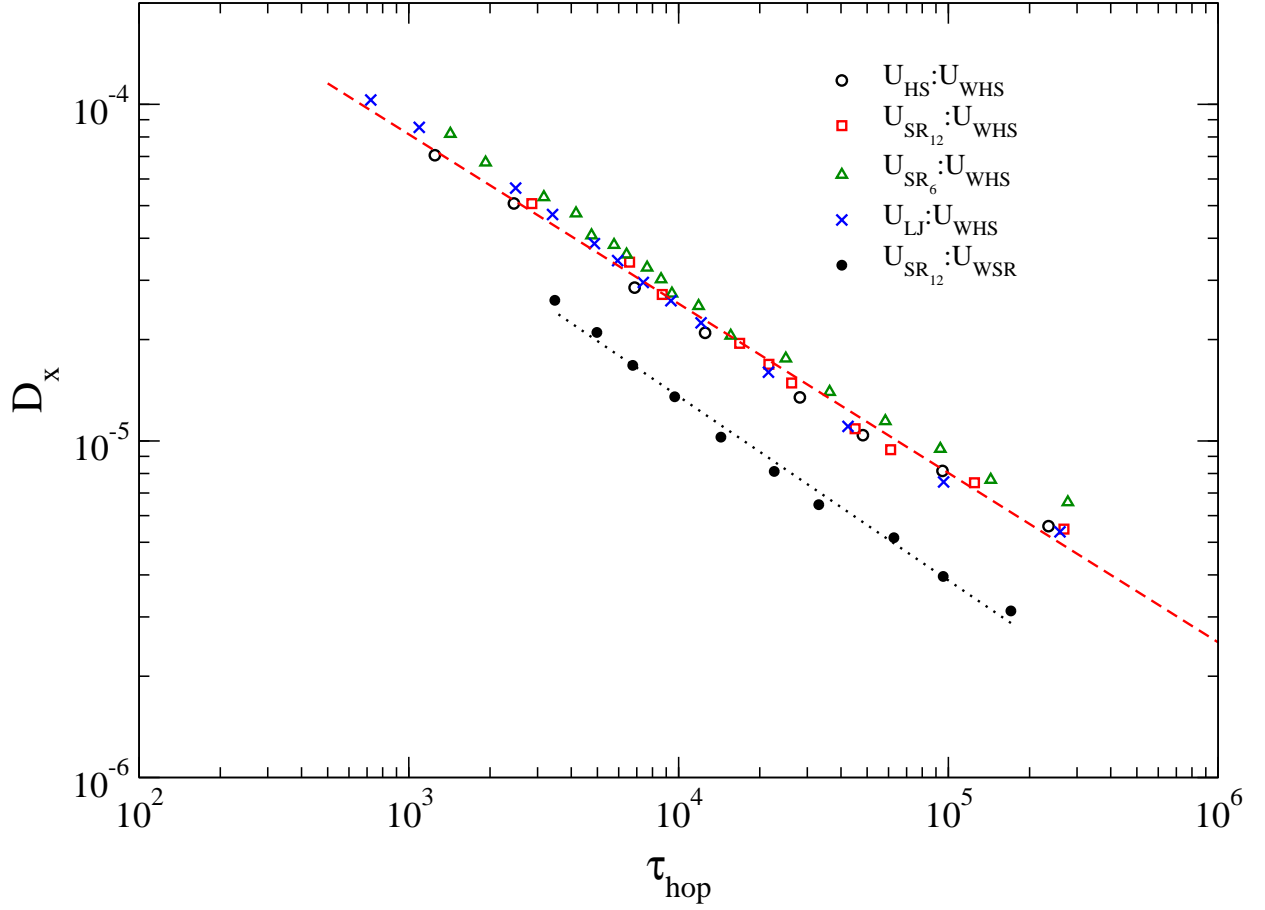}
\caption{The diffusion coefficient, $D_x$, as a function of hopping time, $\tau_{hop}$, for the different systems studied. The dashed line represents the best power law to the $U_{HS}:U_{WHS}$ data and has a slope of -0.5 (i.e., $D_x\sim\tau_{hop}^{-0.5}$). The dotted line is the best fit power law to the $U_{SR_{12}}:U_{WSR}$ data and has a slope of -0.54 (i.e., $D_x\sim\tau_{hop}^{-0.54}$). }
 \label{MSD_hop}
\end{figure}

To test the predictions of Eq.~\ref{eq:dxhop}, we plot $D_x$  as a function of $\tau_{hop}$ in log--log scale (Figure~\ref{MSD_hop}). The dashed line represents the best fit power law to the data for the $U_{HS}:U_{WHS}$ system and has a slope of 0.50, which is consistent with the theoretical prediction by Mon and Percus~\cite{Mon:2002p4642}. The data for the soft particles systems with hard walls all fall along the same line and individual fits to the data yield slopes in the range of 0.49--0.52. The data for the $U_{SR_{12}}:U_{WSR}$ system are displaced to lower values of $D_x$ and the power law exponent is 0.54. These results strongly suggest that Eq.~\ref{eq:dxhop} is generally true for highly confined systems in the SFD--normal diffusion crossover regime.

\begin{figure}[ht]
\includegraphics[width=6.50in]{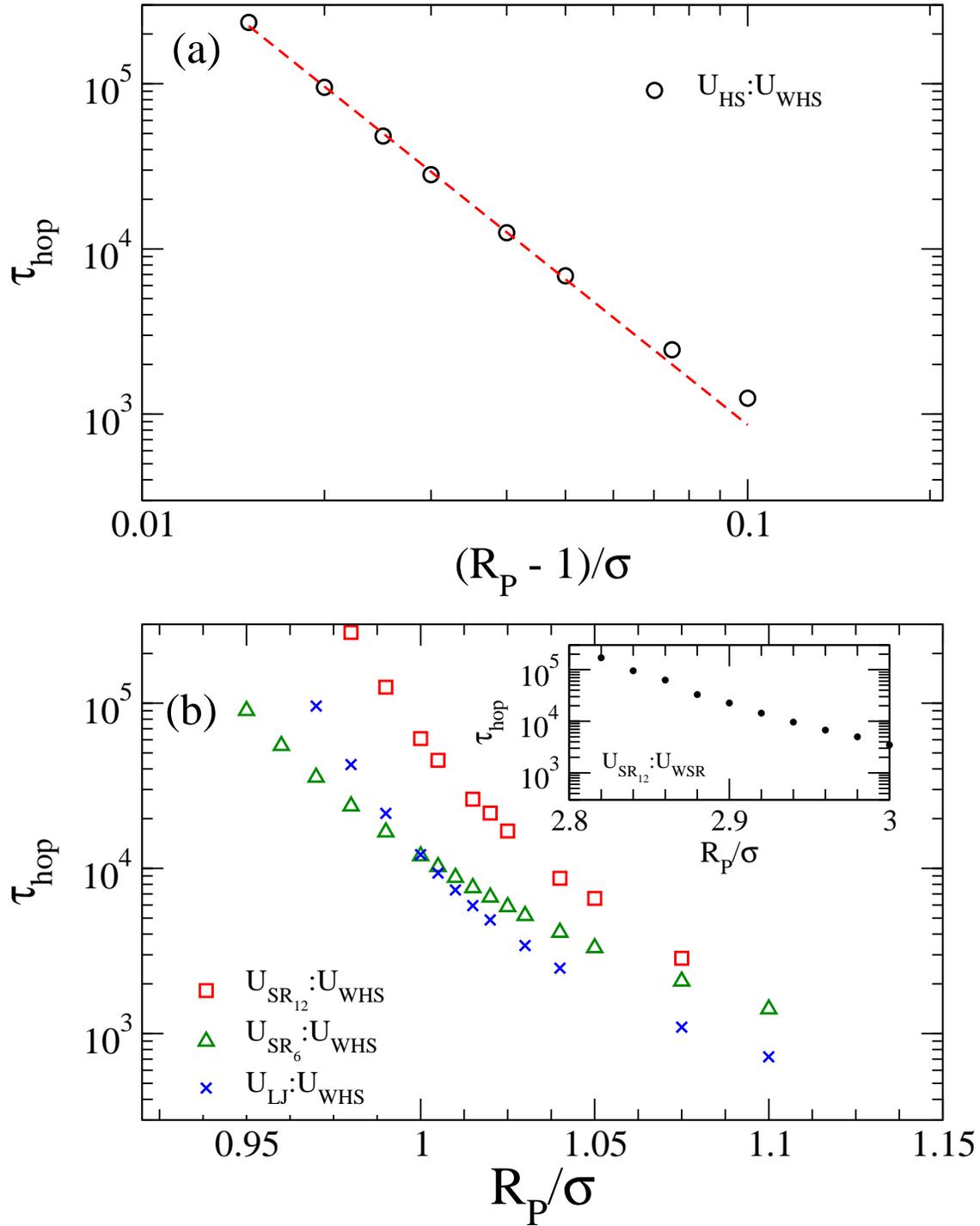}
\caption{(a) Log-Log plot of $\tau_{hop}$, as a function of reduced pore size, $(R_p-1)/\sigma$ for the $U_{HS}:U_{WHS}$ system. The dashed line is the power law fit of Eq.~\ref{eq:hoprp} and has a slope = $-3$. (b) Log-Linear plot of $\tau_{hop}$ as a function $R_p/\sigma$ for the soft particle systems in a hard wall pore. Inset: The same plot for the soft particle-soft wall system.}
 \label{Fig:hop_pore}
\end{figure}

Figure~\ref{Fig:hop_pore}(a) shows that $\tau_{hop}$ for the $U_{HS}:U_{WHS}$ system follows the power law outlined in Eq.~\ref{eq:hoprp} with $\alpha=3$ as predicted by TST.  A power law is not surprising for the purely hard interaction case because the hopping time must diverge as the pore radius approaches the critical passing threshold value. However, this is not true for the soft potential models, where the particles can always pass in principle, except in the limit $D_p\rightarrow 0$, and we see no evidence of power law behavior for the range of $D_p$ studied here (see Figure~\ref{Fig:hop_pore}(b)). San\'{e} et al.~\cite{Sane2010} did find that $\tau_{hop}$ obeyed Eq.~\ref{eq:hoprp} for a model colloid with a $1/r^{48}$ repulsive potential. Although this is very close to being a hard particle, it suggests that it might be possible to find an effective divergence when measuring the hopping times at pore diameters where hopping events become rare on the time scale of the simulation. Nevertheless, our results show that Eq.~\ref{eq:hoprp} is not general and it would be useful to develop an approach that is more widely applicable but still consistent with TST.

\begin{figure}[ht]
\includegraphics[width=7.00in]{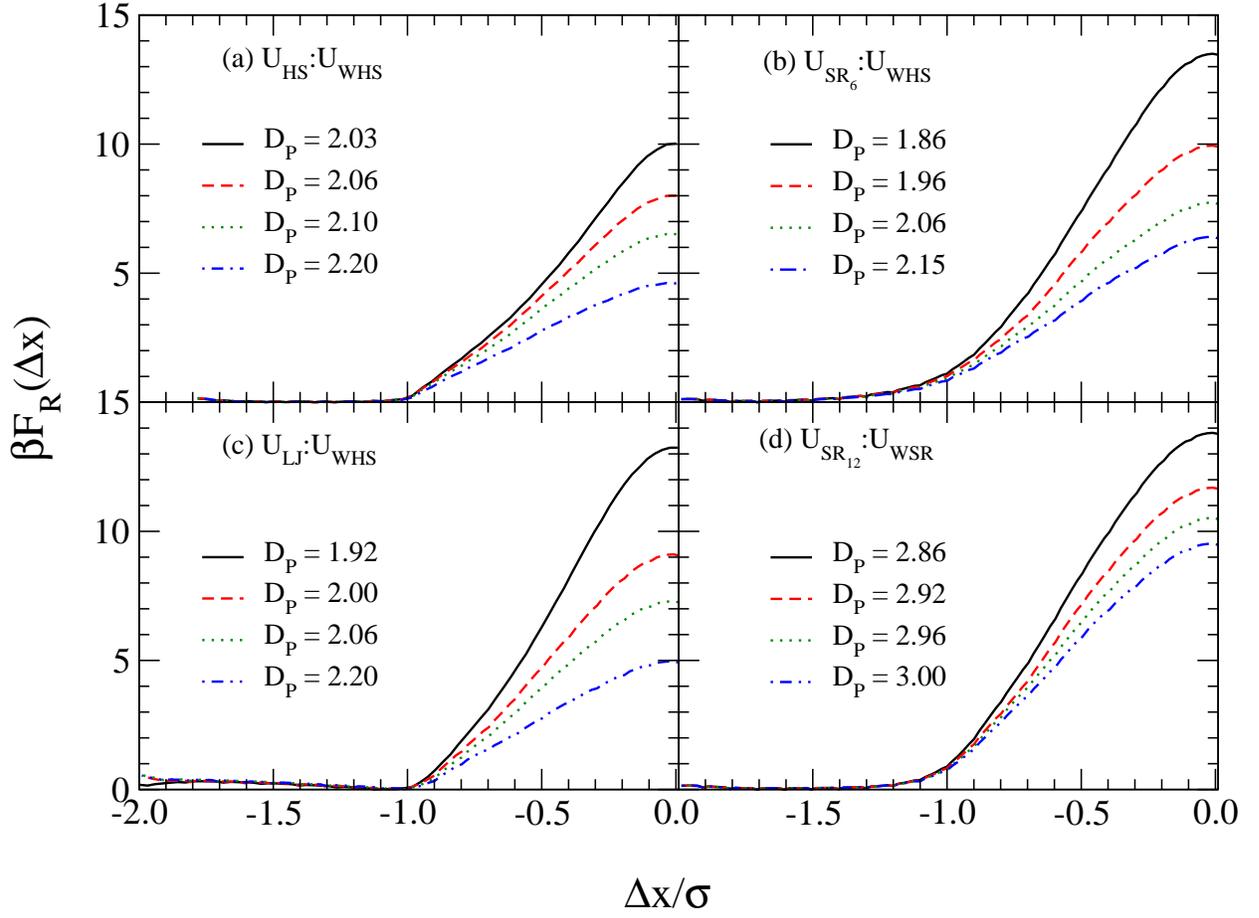}
\caption{ $\beta F_R(\Delta x)$ as a function the distance along the reaction coordinate, $\Delta x/\sigma$, for systems (a) $U_{HS}:U_{WHS}$ (b) $U_{SR_6}:U_{WHS}$ (c) $U_{LJ}:U_{WHS}$ and (d) $U_{SR_{12}}:U_{WSR}$, for a variety of pore diameters.}
 \label{PMF_graphs}
\end{figure}

\begin{figure}[ht]
\includegraphics[width=7.0in]{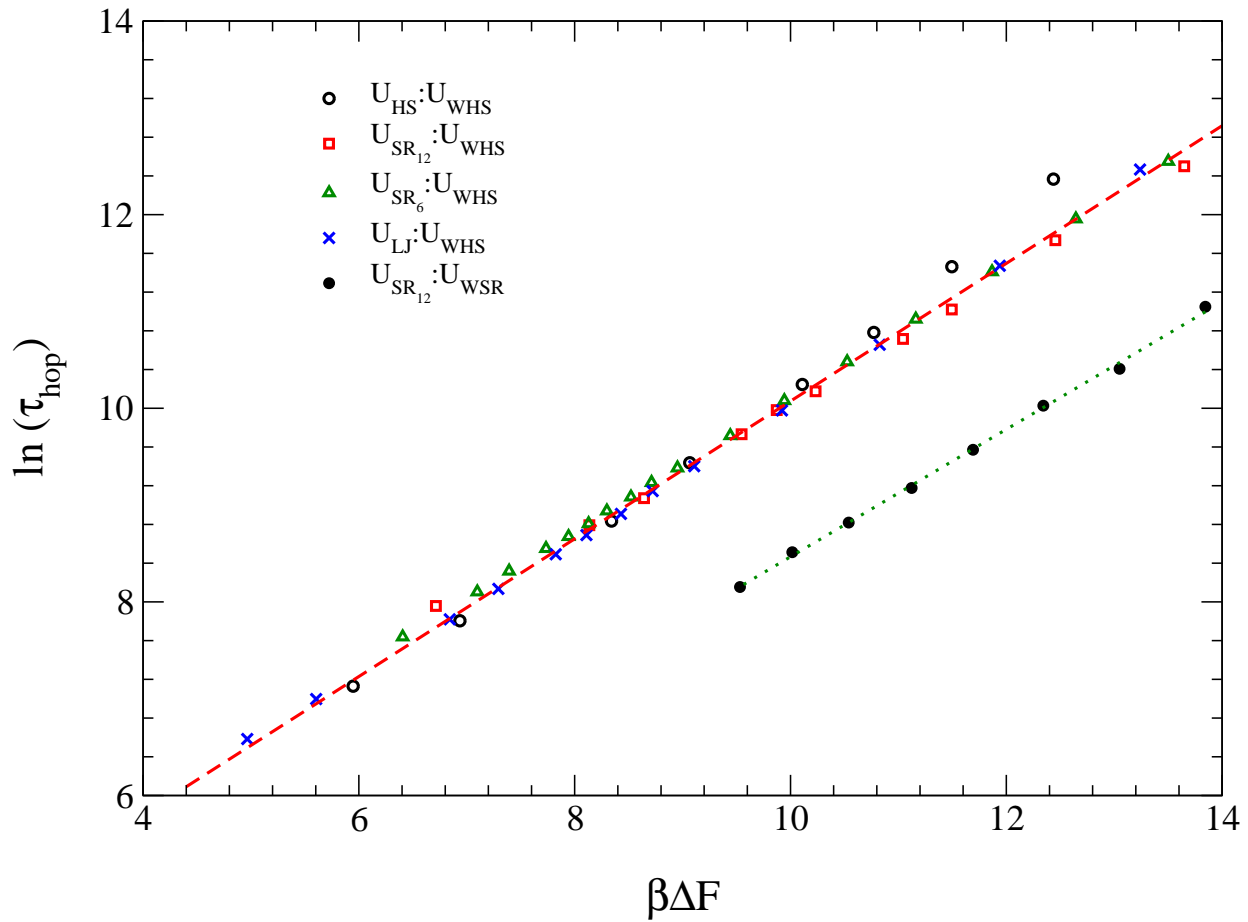}
\caption{$\ln \tau_{hop}$ as a function of $\beta\Delta F$ for all the systems studied. The dashed and dotted represent best linear fits to the data and  have slopes of 0.71 and 0.66, respectively.}
 \label{PMF}
\end{figure}

Figure~\ref{PMF_graphs} shows the free energy profiles for our systems along the two particle reaction coordinate, $\Delta x$, at selected pore diameters. 
(See supplementary material at [URL will be inserted by AIP] for the free energy curve for the $U_{SR_{12}}:U_{WHS}$ model.) The free energy maximum is located at the transition state ($\Delta x=0$), and the barrier height increases as the pore diameter decreases. $\beta F_R(\Delta x)$ decreases to zero at or just beyond $\Delta x=-\sigma$ for systems involving purely repulsive particle-particle interactions ($U_{HS}:U_{WHS}$, $U_{SR_{12}}:U_{WHS}$, $U_{SR_{6}}:U_{WHS}$, $U_{SR_{12}}:U_{WSR}$) but the Lennard--Jones model exhibits a small minimum, reflecting features of the particle-particle interaction potential. The barrier height used in Eq.~\ref{eq:lntaudf} is taken to be $\Delta F=F_R(0)-F_R(-2)$ and Figure~\ref{PMF} shows $\ln (\tau_{hop})$ is linear in  $\Delta F$ for all the systems studied, over a large range of hopping times and pore diameters, indicating that TST provides a useful formalism for calculating the hopping times and subsequently the diffusion coefficients for these highly confined systems. It is also interesting to note that all the systems with the same hard wall interaction potential fall on the same line whereas the soft wall interaction is displaced, consistent with Eq.~\ref{MSD_hop} and probably a consequence of the high degree of confinement induced by the wall potential.

The two particle model used in our the TST approach clearly gives a good description of the hopping time, despite its extreme simplification of the system.
However, it does not give us a quantitative prediction of $\tau_{hop}$. If it did, the slopes of the linear fit curves in Figure \ref{PMF} would be unity. TST requires the free energy be directly related to the probability of finding the system at the top of the barrier. The $\Delta F$ used here gives us the probability of being in the transition state relative to the probability of the particles being separated by $2\sigma$. This differs from the required probability by a normalization factor. A more rigorous treatment of the three particle cage and the integration of the degrees of freedom associated with the $N-3$ particles outside the cage is needed before quantitative predictions can be achieved. The kinetic prefactor that arises in TST must also be determined. In particular, the effects of barrier recrossing would need to be considered because particles that immediately return to their original cage should not contribute diffusive hopping process. A significant amount of recrossing would lead to longer hopping times and lower diffusion constants.

\section{Conclusions}
Highly confined fluids, in a broad range of natural phenomena and engineering applications, exhibit a range of unusual dynamical properties, including anomalous diffusion caused by single file geometry. Our work shows that the hopping time approach, combined with transition state theory, offers a useful and broadly applicable framework for understanding the dynamics of fluids in the SFD--normal diffusion crossover regime. It also suggests that this approach could ultimately lead to quantitative predictions of hopping times and diffusion coefficients for these system.

\acknowledgments
This work was supported by the Natural Sciences and Engineering Research Council (NSERC) of Canada through the Discovery and Engage Grants programs and POS Bio-Sciences.  All computations were performed using computing resources provided by WestGrid and Compute/Calcul Canada.

%\begin{thebibliography}{10}
%\bibliographystyle{jcp}
%\bibliographystyle{phys_chem}
%\bibliography{SFD_ref2}
%\end{thebibliography}
%

\newpage

\noindent Table 1. Selected pore diameter range for each interacting potential and the linear density range obtained from the constant pressure simulation at  $P_L\sigma^3/k_BT$ = 0.4.\\

\noindent Fig 1. The MSD as a function of time, $t$, for systems (a) $U_{HS}:U_{WHS}$ (b) $U_{SR_6}:U_{WHS}$ (c) $U_{LJ}:U_{WHS}$ and (d) $U_{SR_{12}}:U_{WSR}$ for a range of pore diameters. The dashed and dotted lines indicate line slopes  for $\sim t$ and $\sim t^{1/2}$, respectively, for comparison.\\

\noindent Fig 2. The average hopping time, $\tau_{hop}$, as a function of time, $t$, for the narrowest pore diameter studied for each system.\\

\noindent Fig 3. The diffusion coefficient, $D_x$, as a function of hopping time, $\tau_{hop}$, for the different systems studied. The dashed line represents the best power law to the $U_{HS}:U_{WHS}$ data and has a slope of -0.5 (i.e., $D_x\sim\tau_{hop}^{-0.5}$). The dotted line is the best fit power law to the $U_{SR_{12}}:U_{WSR}$ data and has a slope of -0.54 (i.e., $D_x\sim\tau_{hop}^{-0.54}$).\\

\noindent Fig 4. (a) Log-Log plot of $\tau_{hop}$, as a function of reduced pore size, $(R_p-1)/\sigma$ for the $U_{HS}:U_{WHS}$ system. The dashed line is the power law fit of Eq.~\ref{eq:hoprp} and has a slope = $-3$. (b) Log-Linear plot of $\tau_{hop}$ as a function $R_p/\sigma$ for the soft particle systems in a hard wall pore. Inset: The same plot for the soft particle-soft wall system.\\

\noindent Fig 5. $\beta F_R(\Delta x)$ as a function the distance along the reaction coordinate, $\Delta x/\sigma$, for systems (a) $U_{HS}:U_{WHS}$ (b) $U_{SR_6}:U_{WHS}$ (c) $U_{LJ}:U_{WHS}$ and (d) $U_{SR_{12}}:U_{WSR}$, for a variety of pore diameters.\\

\noindent Fig 6. $\ln \tau_{hop}$ as a function of $\beta\Delta F$ for all the systems studied. The dashed and dotted represent best linear fits to the data and  have slopes of 0.71 and 0.66, respectively.\\

\newpage
\begin{table}[h]
\caption{Selected pore diameter range for each interacting potential and the linear density range obtained from the constant pressure simulation at  $P_L\sigma^3/k_BT$ = 0.4.}
\begin{center}
\begin{tabular}{l  c  c  }
\hline
  & $D_P/\sigma$ & $\rho_L\sigma$ \\
  \hline
$U_{HS}:U_{WHS}$ &  2.03 -- 2.20 &  0.42 -- 0.56\\
$U_{SR_{12}}:U_{WHS}$ &  1.96 -- 2.15 &  0.39 -- 0.48\\  
$U_{SR_{6}}:U_{WHS}$ & 1.86 -- 2.15  &  0.34 -- 0.46 \\
$U_{LJ}:U_{WHS}$ & 1.92 -- 2.20  &  0.40 -- 0.57 \\
$U_{SR_{12}}:U_{WSR}$ & 2.82 -- 3.00  &  0.57 -- 0.66\\
\hline
\end{tabular}
\end{center}
\label{Tab:density}
\end{table}

% figures
\newpage

\end{document}